\documentclass[11pt,twoside]{article}

\usepackage{asp2006}
\usepackage{epsf}
\usepackage{psfig}
\usepackage{lscape}
\usepackage{graphicx}

\begin{document}
\title{Coherent lateral motion of Penumbral Filaments during X-class Flare of 13 December 2006}   
\author{Sanjay Gosain, P. Venkatakrishnan and Sanjiv Kumar Tiwari}   
\affil{Udaipur Solar Observatory, P. Box. 198, Dewali, Udaipur 313001, Rajasthan, INDIA}    

\begin{abstract} 
The high-resolution pictures of the solar photosphere from
space based 50 cm Solar Optical Telescope (SOT) onboard {\it
Hinode} spacecraft, are now routinely observed. Such images of
a $\delta$-sunspot in NOAA 10930
 were obtained by {\it Hinode} during 13 December 2006 while a X-class flare occurred in this active
region. Two bright ribbons were visible even in white light and
G-band images apart from chromospheric Ca II H images. We
register the sunspot globally using cross-correlation technique
and analyse local effects during flare interval. We find that
during flare the penumbral filaments show lateral motion. Also,
we locate two patches, one in either polarity, which show
converging motion towards the polarity inversion line (PIL). In
Ca II H images we find kernel with  pre-flare brightening which
lie along the PIL.
\end{abstract}

\section{Introduction}   
The flares in the solar corona are believed to be due to sudden
restructuring of the stressed magnetic field. The energy is
released in the form of thermal as well as non-thermal
radiation and energetic charged particles. In powerful flares
the energetic particles can penetrate the dense chromosphere to
reach down to the photosphere where they heat-up the
photosphere leading to white-light flares. It has been observed
that  photospheric changes are accompanied during these highly
energetic events in the form of: (i) change in morphology, (ii)
change in magnetic flux, (iii) change in magnetic shear angle
(the angle between observed field azimuth and potential field
azimuth) and (iv) proper motion.

Here we focus on the local changes, i.e., changes seen in
small-scale features like penumbral filaments during flares.
Such studies require seeing-free high-resolution observations
at a high-cadence. This is possible with the 50 cm Solar
Optical Telescope (SOT) onboard {\it Hinode} spacecraft
\citep{Kosugi2007,Tsuneta2008,Ichimoto2008,Suematsu2008}. Here,
we present the observations of a $\delta$-sunspot in active
region NOAA 10930 during a X-class flare  on 13 December 2006
at 02:20 UT by {\it Hinode}. The two ribbons could be seen in
G-band and Fe I 630.2 nm Stokes-I and V images
\citep{Isobe2007}. Earlier, we had reported the lateral motion
of penumbral filaments during the flare interval
\citep{Gosain2009}. Here, we present the converging motion of
the two patches, one in either polarity, located on either side
of the PIL. Also, the pre-flare brightening  in kernels located
along the PIL  as seen in Ca II H line, is discussed.

\section{Observations and Data Analysis}
The X-class flare of 13 December 2006 occurred in an active
region numbered NOAA 10930 during 02:20 UT. The region
consisted of a  $\delta$-sunspot with almost N-S orientation of
the bipole. The high-resolution filtergrams in G-band (430.5
nm), Fe I 630.2 nm and Ca II H (396.8 nm) wavelengths were
obtained by Filtergraph (FG) instrument onboard {\it Hinode}
Solar Optical Telescope (SOT). The images were sampled
spatially with 0.1 arc-sec per pixel and temporally with one
image every two minutes.  The filtergrams were calibrated for
dark current, flat field and bad pixels using the standard
SolarSoft IDL libraries. A time sequence of filtergrams was
selected between 02:00 and 03:00 UT for analysis. The images
were aligned globally by choosing a large field-of-view for
registration. The registration procedure for the sunspot in
Hinode filtergrams has been described elsewhere in detail
\citep{Gosain2009}.

\begin{figure}[!ht]
\begin{center}
\includegraphics[scale=0.30,keepaspectratio=true,trim=300 55 0 35]{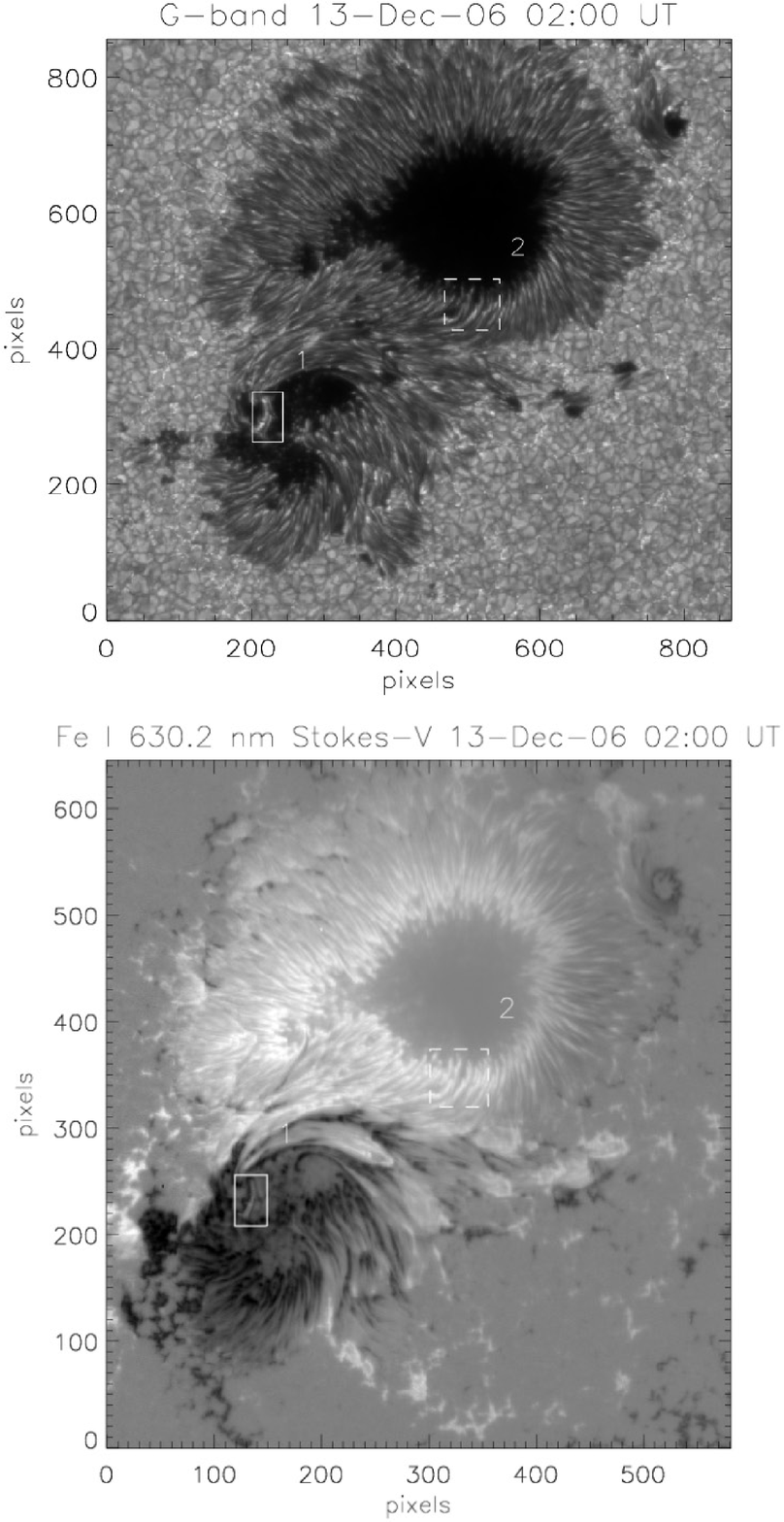}\includegraphics[scale=0.7,keepaspectratio=true,trim=10 0 100 0]{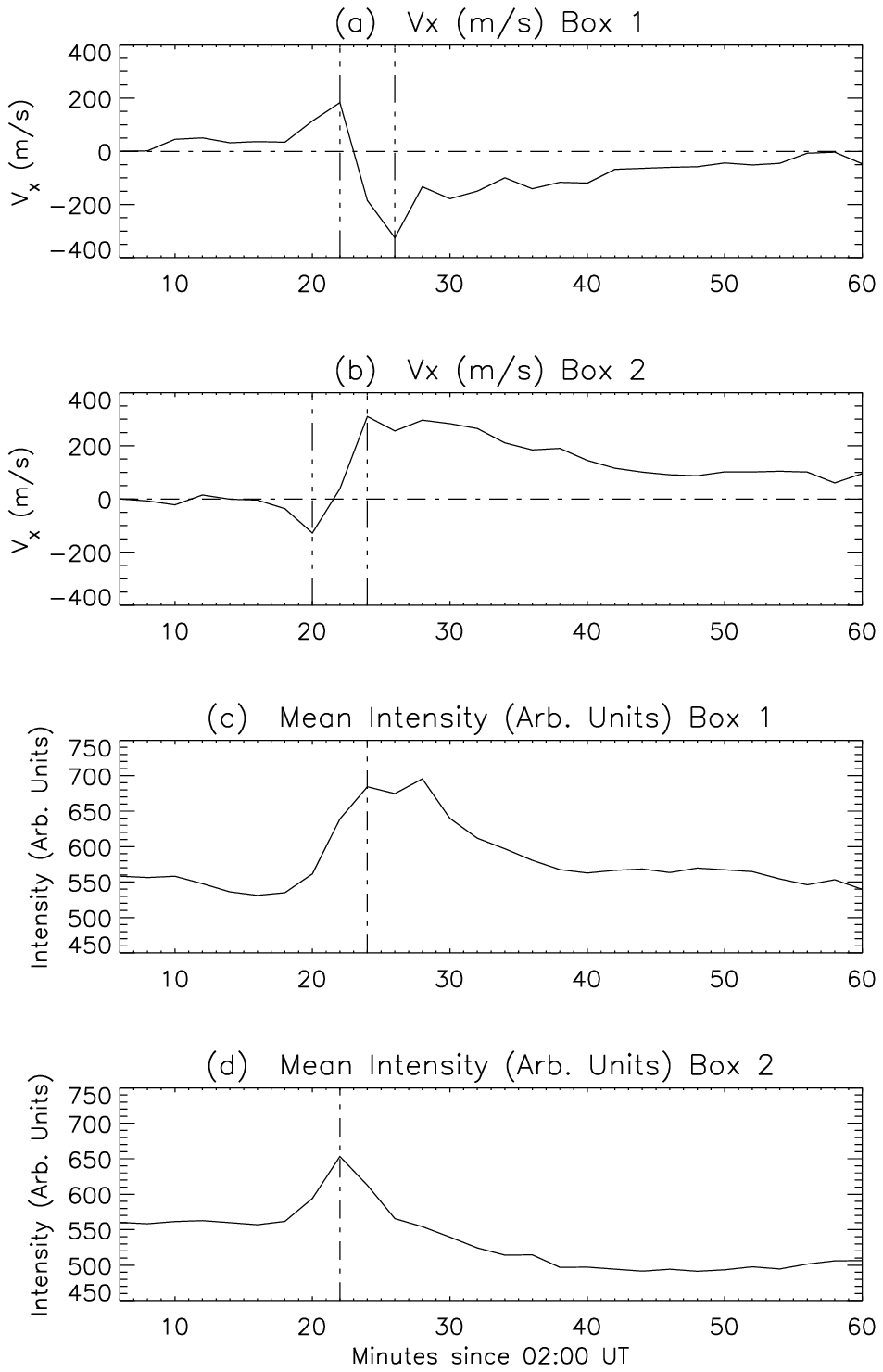}
\end{center}
\caption{The top and bottom panels on the left show the G-band  intensity and Fe I Stokes-V images, respectively.
Boxes `1' and `2' mark the location of the two patches where the velocity $V_x$ and G-band intensity are tracked during the flare interval.
These are plotted in panels (a)-(d) on the right. }\label{fig1}
\end{figure}

\section{Results}
\subsection{Converging Motion of Opposite Polarity Pacthes}
The figure 1 shows the G-band filtergram of the
$\delta$-sunspot taken during 13 December 2006 at 02:00 UT. The
two boxes marked `1' and `2' correspond to patches located in
the two spots of opposite polarity. Within the two patches we
tracked  the relative shifts between two subsequent frames
during the interval 02:06 to 03:00 UT. This shift $\Delta x$
within time interval $\Delta t$ gives velocity $V_x$.  The
panels (a) and (b) of figure 2, show the time profile of $V_x$
corresponding to the patches within the boxes `1' and `2'.

 It may be noticed that : (i) First signatures of motion begin at 02:18 UT which is
  about two minutes earlier than the peak of the flare seen in microwave observations of \cite{Zhang2009}.
 (ii)  For both patches `1' and `2' there are two phases of motion, initially
in one direction and after about four minutes in the opposite
direction. The peak velocity of these two phases of motions are
represented by two vertical lines which are separated by about
four minutes. (iii) The motion in the two boxes are in opposite
direction in both phases. Initially, the motion in two boxes is
away from each other and later it is towards each other, like a
converging motion. (iv) The motion in box `2', which is closer
to the neutral line,  starts earlier than motion in box `1' by
about four minutes. (v) The second phase of motion, i.e.,
converging motion, continues for long duration lasting more
than 40 minutes.

The panels (c) and (d) of figure 2, show the mean G-band
intensity within the two boxes. The two ribbons during the
flare are visible in G-band images. Therefore, the enhancement
in G-band intensity (marked by vertical lines) in panels (c)
and (d) of figure 2, correspond to the instant when the ribbons
move across these boxes. Here also, we notice that the
intensity in box `2', which is located closer to the neutral
line, peaks four minutes earlier than intensity in box `1',
which may explain why motion in box `2' starts earlier than
motion in box `1' by about four minutes.

\subsection{Pre-flare Brightening along PIL in Ca II H}
The figure 3 shows the Ca II H filtergrams of the sunspot
during 13 December 2006 at 02:04, 02:08, 02:14 and 02:20 UT. We
notice that although the flare onset time is 02:20 UT according
to microwave flux observations, initial brightenings can be
noticed in Ca II H filtergrams as early as 02:04 UT in panel
(a). The location of this brightening is marked by a
rectangular box. Further, it may be noticed that the
brightening is located along the neutral line, and in
subsequent panels (a)-(c) the length of the brightening
increases along the neutral line. Finally, the flare develops
into a two ribbon flare as seen in panel (d).

\begin{figure}[!ht]
\begin{center}
\includegraphics[scale=0.62,keepaspectratio=true,trim=0 45 0 15]{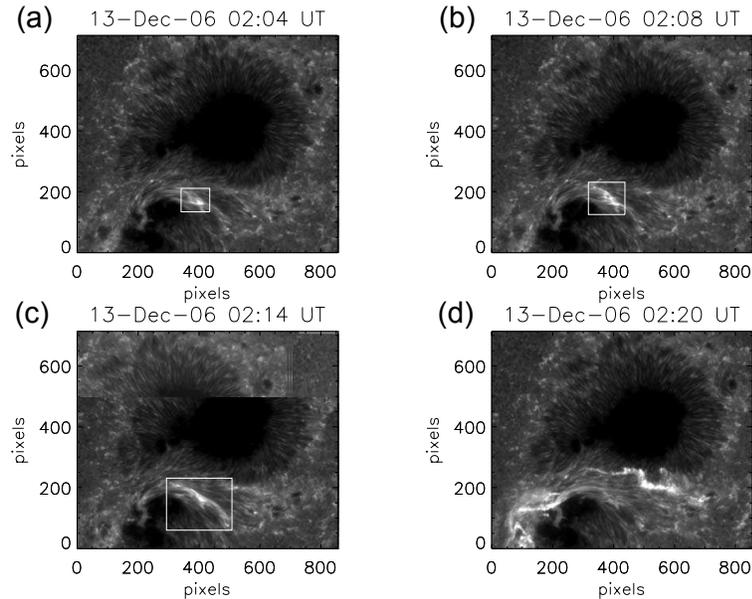}
\end{center}
\caption{The panels (a)-(c) show the Ca II H filtergrams during pre-flare phase. The boxes mark the location of pre-flare
brightening along neutral line. The panel (d) shows the filtergram at 02:20 UT.  }\label{fig3}
\end{figure}

\section{Discussion and Conclusions}
We have studied the evolution of small scale features in a
flaring sunspot using high resolution space based observations.
The fine structure of the sunspot penumbra as seen in high
resolution G-band images is believed to outline the magnetic
field lines. Using the penumbral filaments as a proxy for
magnetic field structure of the sunspot we study the changes in
the structure during X-class flare of 13 December 2006. A two
phased motion is seen in  patches of either poalrity, i.e.,
boxes `1' and `2'. First phase of motion lasts for about four
minutes, directed away from the neutral line, and another phase
of motion  lasts for more than 40 minutes, directed towards
neutral line, i.e., converging.

Further, we notice that in Ca II H images the pre-flare
brightening is clearly visible in elongated kernels, located
along the PIL. This brightening is seen as early as 16 minutes
prior to the flare onset.

To understand the changes in magnetic field configuration
during flares one requires high-cadence vector magnetograms
obtained with high-resolution which are expected from upcoming
space missions like Helioseismic and Magnetic Imager (HMI) and
Solar Orbiter.

\acknowledgements 
{\it Hinode} is a Japanese mission developed and launched by
ISAS/JAXA, with NAOJ as a domestic partner and NASA and STFC
(UK) as international partners. It is operated by these
agencies in co-operation with ESA and NSC (Norway).

\end{document}